\documentclass[NewProceedings,SingleSpace,letterpaper]{ascelike-new}
\usepackage[utf8]{inputenc}
\usepackage[T1]{fontenc}
\usepackage{lmodern}
\usepackage{graphicx}
\usepackage[style=base,figurename=Fig.,labelfont=bf,labelsep=period]{caption}
\usepackage{subcaption}
\usepackage{amsmath}
\usepackage{amsfonts}
\usepackage{amssymb}
\usepackage{amsbsy}
\usepackage{colortbl}
\usepackage{float}
\usepackage{natbib}
\usepackage{newtxtext,newtxmath}
\usepackage[colorlinks=true,citecolor=red,linkcolor=black]{hyperref}

\NameTag{Roy, \today}
\begin{document}

\title{PRISM: AN AGENTIC MULTI-MODEL ARCHITECTURE FOR PROACTIVE SAFETY IN AUTONOMOUS TRANSPORTATION SYSTEMS}

\author[1]{Joyjit Roy}
\author[2]{Samaresh Kumar Singh}
\author[3]{Sushanta Das, Ph.D.}
 
\affil[1]{Independent Researcher, IEEE Senior Member, Austin, Texas, USA.
Email: joyjit.roy.tech@gmail.com}
\affil[2]{Independent Researcher, IEEE Senior Member, Leander, Texas, USA.
Email: ssam3003@gmail.com}
\affil[3]{American Center for Mobility, Ypsilanti, Michigan, USA.
Email: sushanta.das@acmwillowrun.org}

\maketitle

\begin{abstract}
Autonomous and intelligent transportation systems operate within complex urban environments, where safety relies on interactions among vehicle behavior, environmental conditions, and vulnerable road users (VRUs), including pedestrians and cyclists. Most advanced driver assistance systems (ADAS) primarily employ reactive mechanisms, including collision detection and emergency braking, that activate only after hazards have emerged. The rise in VRU fatalities in the United States over the past decade underscores a critical limitation of vehicle-centric safety strategies.

This study introduces \textbf{PRISM (Proactive Risk Intelligence and Safety Management)}, an agentic multi-model safety architecture that transitions from reactive crash avoidance to proactive, continuous risk management. The proposed framework employs inverse crash-probability modeling to convert binary crash classifiers trained on national crash statistics into dynamic, interpretable safety scores. Three specialized models addressing trajectory kinematics, weather and road-surface assessment, and VRU interaction prediction operate concurrently and are coordinated by an adaptive reasoning layer that incorporates reinforcement learning, contextual memory, and feature-level attribution to enhance explainability. The agentic layer integrates these signals to provide graduated safety interventions across four tiers, ranging from silent monitoring to emergency alerts.

Unlike rule-based systems with static thresholds, the proposed architecture dynamically adjusts safety parameters in real time according to the driving context. The framework was validated using multiple large-scale naturalistic driving datasets without dataset-specific retraining, demonstrating consistent cross-domain behavior. On naturalistic validation data, the system yielded a mean safety score of 68 out of 100, classified 77.6\% of scenarios as advisory, and flagged a near-miss rate of 3.8\%, with approximately 11\% of scenarios escalating to intervention or emergency response. Feature attribution consistently highlighted trajectory risk and VRU proximity as the primary safety factors. This approach provides a unified and interpretable framework for proactive transportation safety, emphasizing the reduction of VRU risk in dense urban environments.
\end{abstract}
\section{Practical Applications}
\label{sec:practical}

This research offers a practical framework to improve safety in automated and assisted driving in busy urban environments. Instead of reacting only after danger arises, the system continuously assesses the risk of the current driving context and responds with interventions calibrated to that risk, from passive monitoring to urgent alerts. For vehicle manufacturers and advanced driver-assistance system (ADAS) developers, this graduated strategy minimizes unnecessary warnings while ensuring hazardous situations receive prompt attention. For fleet operators, it aggregates risk across extensive trip data to identify dangerous routes, peak risk periods, and unsafe driving patterns without requiring a crash. For city and infrastructure planners, the same metrics, especially those involving pedestrian and cyclist safety, can guide crosswalk placement, signal timing, and school zone protections. Each intervention includes a clear explanation of its contributing factors, supporting public trust, regulatory oversight, and independent evaluation of automated safety systems.
\section{Introduction}
\label{sec:intro}

Autonomous and intelligent transportation systems are increasingly used in dense urban areas, where safety depends on the continuous interaction among vehicle behavior, environmental conditions, and VRUs, including pedestrians and cyclists. Although automated driving can reduce crashes caused by human error, maintaining safety in mixed traffic remains a significant challenge. VRU fatalities in the United States have reached their highest levels in decades despite advances in vehicle safety technology \citep{nhtsa_fatality, ghsa_ped}, underscoring a persistent disconnect between vehicle-centric safety design and the complexities of shared urban mobility.

Most advanced driver assistance systems (ADAS) in production are fundamentally reactive. Features like automatic emergency braking and forward collision warning activate only after a hazard has materialized, leaving a narrow margin for intervention, and they rely on fixed thresholds that do not adapt to context \citep{adas_review}. Consequently, a maneuver deemed safe on an open highway is treated the same as one performed near a crowded crosswalk at night, leading to unaddressed risks in complex environments and nuisance alerts in benign ones.

An alternative approach conceptualizes safety as a continuous, predictive measure rather than a discrete event trigger. In our prior work, we introduced an inverse crash-probability method that transforms a binary crash classifier, trained on large-scale national crash records, into a calibrated safety score \citep{safedriveriq}. That study found that environmental context and the co-occurrence of multiple risk factors, rather than driver aggression alone, are the primary determinants of crash risk. However, it was less responsive to dynamic, scene-level indicators of imminent VRU risk, including relative trajectories, closing speeds, and time to collision.

Industry-scale mobility analytics highlight the need for context-aware safety architectures. The Arity Annual Driving Behavior Report \citep{arity2025driving}, drawing on more than 45 million anonymized U.S. drivers and two trillion miles of trips, identifies hard braking, sudden acceleration, high-speed mileage, and distracted driving as key behavioral risk indicators. The report sets thresholds of 8 mph per second for both hard braking and sudden acceleration. It also demonstrates that driving behavior varies substantially with commuting patterns, seasonal conditions, weather events, and emergency situations. For example, during the 2025 Los Angeles wildfire evacuation, hard braking increased by 25.3\% and speed decreased by 17\%. These findings confirm that static threshold-based systems cannot address the compound, context-dependent nature of real-world crash risk, motivating the agentic, memory-enabled architecture presented here.

This paper builds on that foundation by introducing an agentic multi-model safety architecture that transitions from reactive crash avoidance to proactive, continuous risk management. Three specialized models operate in parallel to assess environmental risk, trajectory kinematics, and VRU interaction. An agentic reasoning layer fuses their outputs through reinforcement learning (RL), contextual memory, and feature-level attribution. This approach enables a graduated set of interventions, from silent monitoring to emergency alerts, each with an explanation of contributing factors. The main contributions are:

\begin{itemize}
\item \textbf{An agentic multi-model architecture} that unifies environmental, kinematic, and VRU-interaction risk into a single interpretable safety score with a graduated, four-tier intervention policy.
\item \textbf{A reasoning layer} combining RL, contextual memory, and feature attribution for context-adaptive yet explainable decisions.
\item \textbf{Cross-dataset validation} on multiple naturalistic driving datasets without dataset-specific retraining, identifying trajectory risk and VRU proximity as the dominant safety factors.
\end{itemize}
\section{Related Work}
\label{sec:related}

Prior research on automotive safety spans three broad directions: statistical crash-risk modeling, trajectory and VRU interaction prediction, and learning-based or explainable decision systems. Table~\ref{tab:related} summarizes key approaches and the safety dimensions each addresses. Statistical models derived from national crash databases assess environmental and contextual risk, but only at the aggregate trip level, not in real time \citep{safedriveriq}. Trajectory and interaction models, including the Social Force Model \citep{helbing1995} and recurrent predictors such as Social LSTM \citep{alahi2016}, capture dynamic motion and VRU proximity but lack a calibrated, scenario-level safety measure. Learning-based controllers \citep{kiran2021} and post hoc explainability methods \citep{lundberg2017} contribute to adaptive decision-making and interpretability, but are seldom integrated into a single architecture that delivers a continuous, explainable safety score with graduated intervention outputs.

\begin{table}[ht]
\centering
\caption{Comparison of representative safety approaches. Env, environmental risk; Kin, trajectory kinematics; VRU, VRU interaction; Cont, continuous score; Expl, explainable; Grad, graduated intervention.}
\label{tab:related}
\centering
\footnotesize
\begin{tabular}{lcccccc}
\hline
\textbf{Approach} & \textbf{Env} & \textbf{Kin} & \textbf{VRU} & \textbf{Cont} & \textbf{Expl} & \textbf{Grad} \\
\hline
Crash-statistics models \citep{safedriveriq} & \checkmark & -- & -- & \checkmark & \checkmark & -- \\
Social Force Model \citep{helbing1995} & -- & -- & \checkmark & -- & -- & -- \\
Recurrent trajectory prediction \citep{alahi2016} & -- & \checkmark & \checkmark & -- & -- & -- \\
Reactive ADAS \citep{adas_review} & -- & \checkmark & -- & -- & -- & -- \\
RL-based driving policies \citep{kiran2021} & -- & \checkmark & \checkmark & -- & -- & \checkmark \\
Post-hoc explainability \citep{lundberg2017} & -- & -- & -- & -- & \checkmark & -- \\
\textbf{This work} & \checkmark & \checkmark & \checkmark & \checkmark & \checkmark & \checkmark \\
\hline
\end{tabular}
\end{table}

As Table~\ref{tab:related} shows, previous approaches address only parts of the safety problem. No existing framework combines environmental, kinematic, and VRU-interaction risks into a single, continuous, and explainable score with graduated intervention. This work aims to fill that gap.
\section{PRISM: System Architecture}
\label{sec:architecture}

\begin{figure}[ht]
\centering
\includegraphics[width=\linewidth]{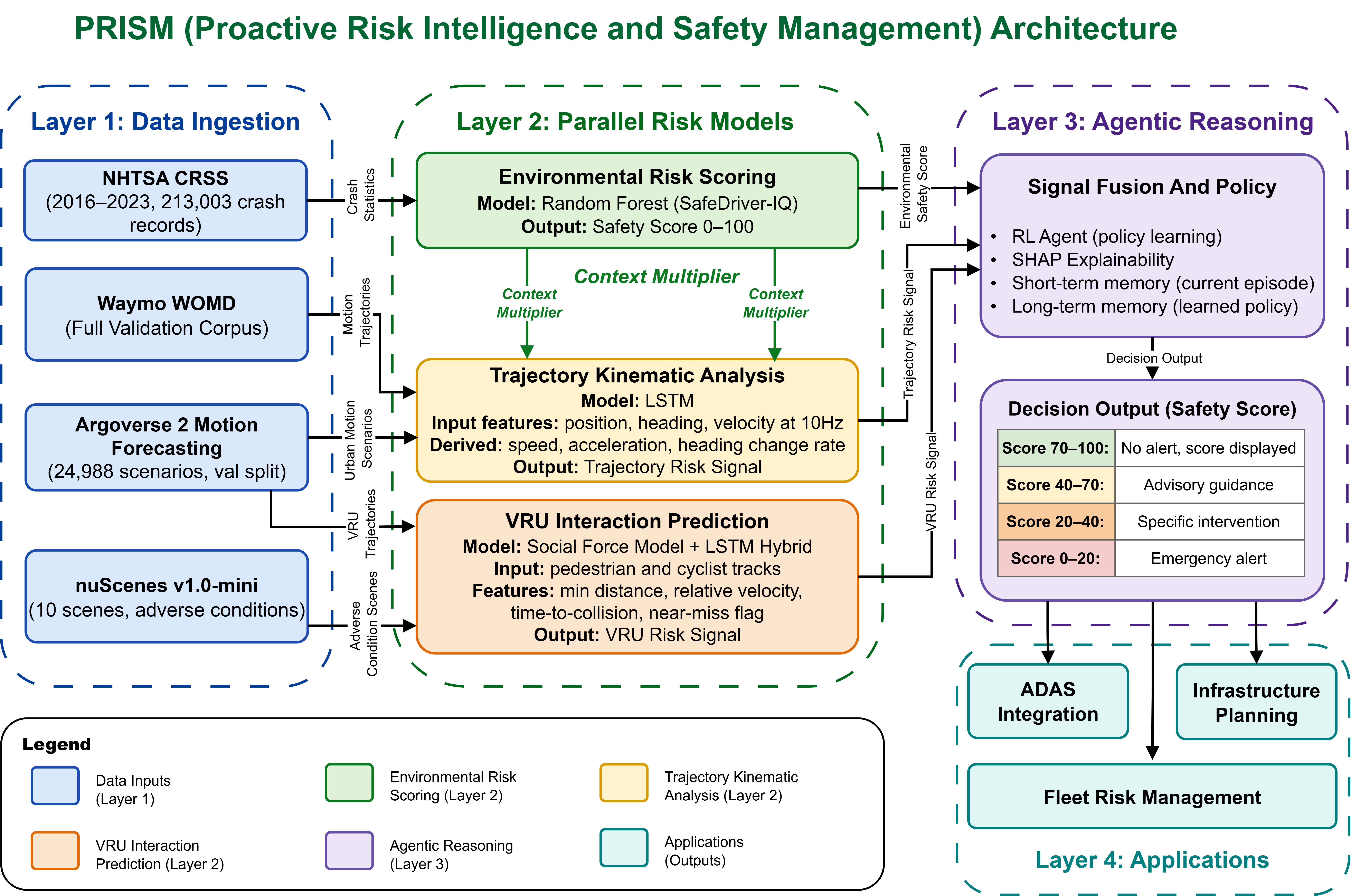}
\caption{Four-layer agentic multi-model architecture. Layer~2 runs three risk models in parallel; Layer~3 fuses them, applies the reinforcement learning policy, and generates an explanation.}
\label{fig:architecture}
\end{figure}

The proposed system is a four-layer architecture that converts raw multi-agent motion data into a continuous, explainable safety score with graduated intervention outputs. Figure~\ref{fig:architecture} illustrates the overall design. Layer~1 ingests and normalizes scenario data from heterogeneous sources. Layer~2 executes three independent risk models in parallel: an environmental model, a trajectory kinematic model, and a VRU interaction model. Layer~3, the agentic reasoning layer, fuses the three risk signals, applies a reinforcement learning (RL) policy to select an intervention tier, and provides a SHAP-based explanation supported by short-term and long-term memory. Layer~4 maps the decision to downstream applications. A key design principle is that all three risk models operate together on every cycle, ensuring that no single failure mode creates a blind spot.

\subsection{Layer 1: Data Ingestion and Normalization}
\label{subsec:layer1}

The three source datasets use incompatible on-disk formats. The ingestion layer converts each dataset into a unified \texttt{DrivingScene} object, which contains \texttt{AgentTrack} records with position, velocity, heading, object type, and validity mask at 10~Hz, along with scene-level night and rain flags. Agent types are normalized to vehicle, pedestrian, or cyclist, with pedestrians and cyclists additionally grouped as VRUs. For Waymo data, where the type field is frequently ambiguous, classification falls back to bounding-box dimensions to ensure no track is omitted.

A parallel transformation prepares data for the environmental model by projecting scene metadata into a CRSS-style feature dictionary. This ensures the frozen Phase~1 random forest receives inputs in its original feature space, eliminating the need for retraining. This two-track normalization enables all three Layer~2 models to share a single ingestion pass over the raw data.

\subsection{Layer 2: Parallel Risk Models}
\label{subsec:layer2}

\subsubsection{Environmental Risk Scoring.}
The environmental model reuses our prior SafeDriver-IQ random forest \citep{safedriveriq} as a frozen context estimator. It produces an inverse-crash safety score $S_{env} \in [0,100]$, which is converted into a normalized environmental risk

\begin{equation}
r_{env} = \mathrm{clip}_{[0,1]}\!\left(\frac{100 - S_{env}}{100}\right),
\label{eq:envrisk}
\end{equation}

and then into a context multiplier

\begin{equation}
m = m_{\min} + (m_{\max} - m_{\min})\, r_{env},
\label{eq:multiplier}
\end{equation}

with $m_{\min}=0.5$ and $m_{\max}=1.5$. A benign environment damps dynamic risk ($m<1$) while a hostile one, such as night or rain, amplifies it ($m>1$). The environmental model, therefore, acts as a multiplier rather than an equal vote, directly addressing the inability of the standalone Phase~1 model to respond to dynamic context.

\subsubsection{Trajectory Kinematic Analysis.}
The trajectory model evaluates each agent's motion quality by calculating speed, longitudinal and lateral acceleration, and yaw rate for each track using smoothed finite differences. The model assigns a per-timestep kinematic risk from normalized exceedances of comfort and aggression thresholds for hard braking, hard acceleration, swerving, and speeding. The track-level risk $r_{traj} \in [0,1]$ reflects the aggregated exceedance over the agent's trajectory. The scene-level trajectory risk is the highest risk observed among all agents.

\subsubsection{VRU Interaction Prediction.}
The VRU model estimates ego-VRU conflict risk using a Social Force Model \citep{helbing1995} for reaction-aware rollout, combined with a recurrent predictor based on Social LSTM \citep{alahi2016}. For each ego-VRU pair, it calculates the closest approach distance and time-to-collision proxy (TTC), then combines these into an interaction risk $r_{vru} \in [0,1]$ that increases as distance and TTC fall below warning thresholds. A near-miss is flagged when the minimum distance drops below 2.0~m for pedestrians or 1.5~m for cyclists.

\subsubsection{Risk Fusion.}
The three signals are fused into a single dynamic risk. The trajectory and VRU risks are first blended with VRU-dominant weights $w_t = 0.5$ and $w_v = 1.0$, reflecting that VRU conflicts are the safety-critical case,

\begin{equation}
r_{base} = \frac{w_t\, r_{traj} + w_v\, r_{vru}}{w_t + w_v}.
\label{eq:base}
\end{equation}

The environmental multiplier then scales this dynamic risk, and the result is mapped to a continuous safety score,

\begin{align}
r_{fused} &= \mathrm{clip}_{[0,1]}\!\left(m\, r_{base}\right), \label{eq:fused} \\
S         &= 100\,(1 - r_{fused}).                              \label{eq:score}
\end{align}

Equations~\eqref{eq:base}-\eqref{eq:score} define the PRISM heuristic baseline; the RL agent in Layer~3 learns the final intervention policy on top of the same state.

\subsection{Layer 3: Agentic Reasoning}
\label{subsec:layer3}

\subsubsection{Reinforcement Learning Policy.}
The reasoning layer encodes each scenario as a fixed eight-dimensional state vector comprising environmental risk, environmental multiplier, trajectory risk, VRU risk, normalized proximity and imminence (TTC) surrogates, and night and rain flags. A deep Q-network \citep{kiran2021} maps this state to one of four intervention tiers (silent, advisory, intervention, emergency) by selecting the action with the highest estimated value,

\begin{equation}
a^\star = \arg\max_{a}\, Q(\mathbf{s}, a).
\label{eq:qpolicy}
\end{equation}

Learning the policy, rather than thresholding the heuristic score, lets the system weigh combinations of risk factors that fixed rules treat independently.

\subsubsection{Explainability and Memory.}
Each decision includes a SHAP attribution \citep{lundberg2017} across the eight state features, highlighting the factors influencing the selected tier. Short-term memory (STM) maintains the current episode for temporal consistency, while long-term memory (LTM) stores representative past states, enabling the agent to recall similar situations and outcomes. Together, these elements provide a human-readable rationale for each intervention.

\subsection{Layer 4: Applications}
\label{subsec:layer4}

The graduated output supports three application domains without retraining. In ADAS integration, the tier sets driver feedback intensity. For fleet risk management, continuous scores are aggregated into route and driver risk profiles. In infrastructure planning, locations with consistently low scores identify high-conflict sites.

\subsection{Graduated Intervention Design}
\label{subsec:intervention}

Instead of a single binary alert, the system emits one of four tiers based on the safety score $S$, as summarized in Table~\ref{tab:tiers}. The bands escalate from silent monitoring in safe conditions to an emergency alert in severe risk, minimizing nuisance alerts while ensuring a strong response when necessary.

\begin{table}[ht]
\centering
\caption{Graduated intervention tiers as a function of the continuous safety score $S$.}
\label{tab:tiers}
\small
\begin{tabular}{lcl}
\hline
\textbf{Tier} & \textbf{Score range} & \textbf{Response} \\
\hline
Silent       & $70 \le S \le 100$ & Monitor only, no driver alert \\
Advisory     & $40 \le S < 70$    & Soft cautionary feedback \\
Intervention & $20 \le S < 40$    & Specific corrective guidance \\
Emergency    & $0 \le S < 20$     & Urgent alert \\
\hline
\end{tabular}
\end{table}
\section{Datasets and Experimental Setup}
\label{sec:datasets}

PRISM was evaluated on three motion datasets without dataset-specific retraining, as summarized in Table~\ref{tab:datasets}:
\begin{itemize}
\item \textbf{NHTSA CRSS (2016--2023)} \citep{nhtsa_crss}: national crash records used to train the Phase~1 model \citep{safedriveriq}.
\item \textbf{nuScenes v1.0-mini} \citep{nuscenes}: 10 scenes from Boston and Singapore explicitly recorded at night and in rain, serving as an adverse-condition benchmark.
\item \textbf{Argoverse~2 Motion Forecasting} \citep{argoverse2}: validation split across 6 U.S.\ cities, enabling city-level risk analysis.
\item \textbf{Waymo Open Motion Dataset} \citep{waymo}: one validation shard containing high-density autonomous vehicle trajectories.
\end{itemize}

\begin{table}[ht]
\centering
\caption{Datasets used for training and validation.}
\label{tab:datasets}
\footnotesize
\begin{tabular}{lcccc}
\hline
\textbf{Dataset} & \textbf{Role} & \textbf{Scale} & \textbf{Hz} & \textbf{Conditions} \\
\hline
NHTSA CRSS 2016--2023 & Env.\ model training & 213,003 records & -- & All U.S.\ crash types \\
nuScenes v1.0-mini & Phase~2 validation & 10 scenes & 2 & Night, rain \\
Argoverse~2 val split & Phase~2 validation & 24,988 scenarios & 10 & 6 U.S.\ cities \\
Waymo WOMD (1 shard) & Phase~2 validation & 286 scenarios & 10 & Mixed urban \\
\hline
\end{tabular}
\end{table}

Evaluation reports the mean safety score, tier distribution (silent, advisory, intervention, emergency), near-miss rate, and SHAP feature attribution across all datasets. Cross-dataset consistency is assessed by comparing tier distributions and top-ranked SHAP features without dataset-specific adjustment.

\section{Results and Discussion}
\label{sec:results}

PRISM was evaluated on 1,296 scenarios across nuScenes v1.0-mini (10 scenes), Argoverse~2 Motion Forecasting (1,000 scenarios), and Waymo Open Motion Dataset (286 scenarios). No dataset-specific retraining was conducted. Table~\ref{tab:summary} summarizes the key performance metrics across all three datasets.

\begin{table}[ht]
\centering
\caption{Cross-dataset performance summary (1,296 scenarios total).}
\label{tab:summary}
\small
\begin{tabular}{lrrrrrr}
\hline
\textbf{Dataset} & \textbf{n} & \textbf{Mean Score} & \textbf{Advisory\%} & \textbf{Emergency\%} & \textbf{Near-Miss Rate} \\
\hline
nuScenes   & 10    & 59.8 & 70.0 & 20.0 & 10.0\% \\
Argoverse~2 & 1,000 & 68.6 & 76.7 &  4.6 &  7.3\% \\
Waymo WOMD & 286   & 68.0 & 77.6 &  4.5 &  3.8\% \\
\hline
\end{tabular}
\end{table}

\subsection{Safety Score Distribution}
\label{subsec:results_scores}

\begin{figure}[ht]
\centering
\includegraphics[width=\linewidth]{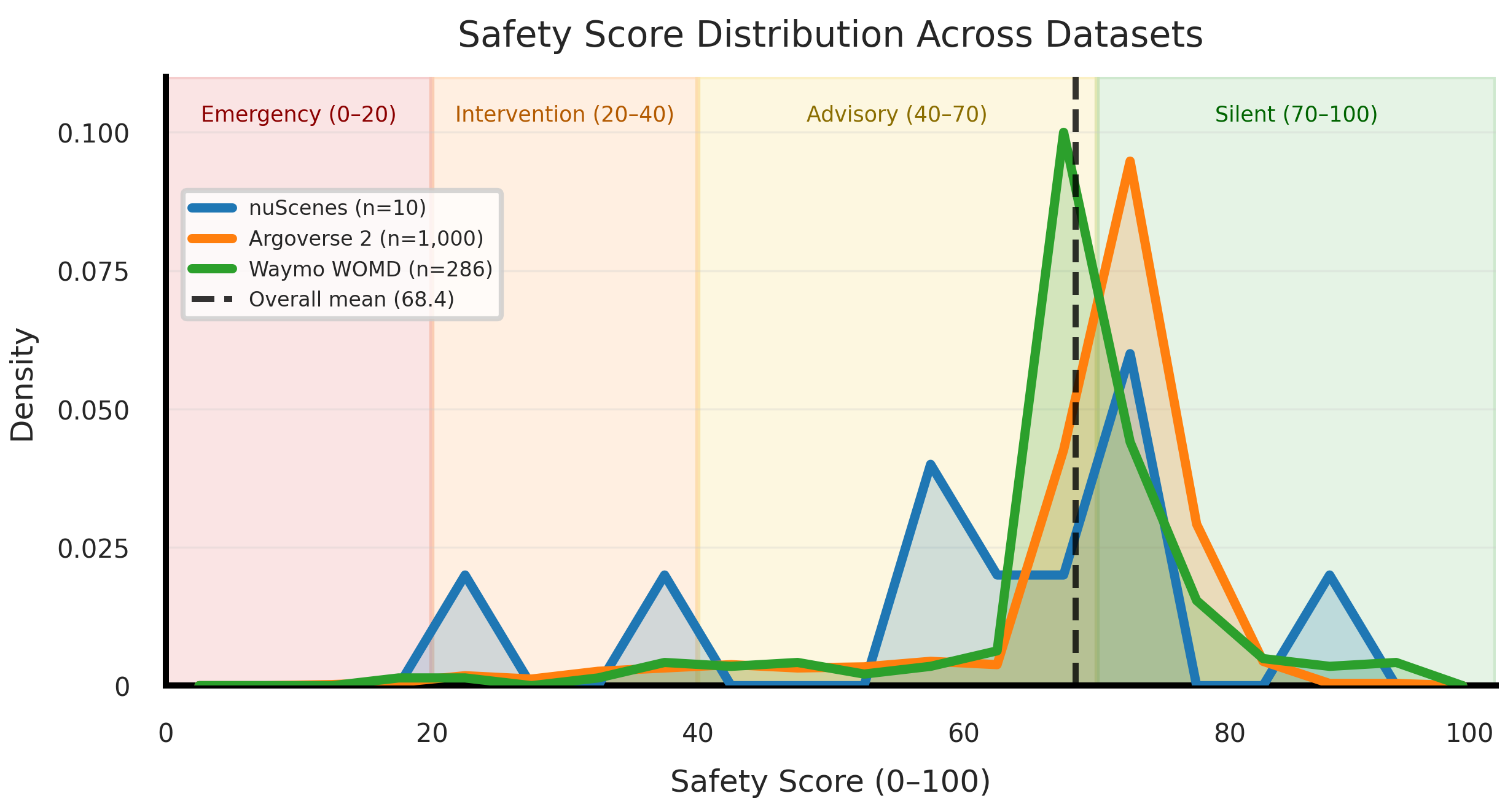}
\caption{Safety score distributions for all three datasets. Shaded regions indicate intervention tiers: emergency (0-20), intervention (20-40), advisory (40-70), and silent (70-100). The dashed line marks the weighted mean (68.4) across all 1,296 scenarios.}
\label{fig:scores}
\end{figure}

Figure~\ref{fig:scores} presents the safety score distributions across all three datasets. Argoverse~2 and Waymo converge to nearly identical means (68.6 and 68.0), despite independent data collection, sensors, and geography, demonstrating cross-domain generalization without retraining. Both distributions are advisory-dominant (40 - 70), consistent with structured urban driving under normal conditions.

nuScenes scores are lower (mean 59.8) and shift toward the intervention and emergency bands, driven by its adverse-condition sampling:
 
\begin{itemize}
\item Three night scenes and one rain scene raise the environmental multiplier to $m = 1.35$, compared to 0.97 for clear daytime, amplifying trajectory and VRU risk signals.
\item The two emergency-tier scenes (scores 35.8 and 23.3) involve high pedestrian density with either a confirmed near-miss or simultaneous night and rain.
\item The single silent-tier scene (score 85.6, three VRUs at 27.7~m) confirms PRISM does not over-escalate in genuinely low-risk conditions.
\end{itemize}

\subsection{Geographic Generalizability}
\label{subsec:results_geo}

Table~\ref{tab:cities} summarizes PRISM performance across six U.S. cities in Argoverse~2. Mean safety scores vary within a narrow 5.3-point range (66.4 - 71.7), confirming generalization across diverse traffic environments without city-specific retraining. Miami (n = 260) has the highest emergency rate (7.3\%), consistent with its dense urban layout, while Palo~Alto (n = 60) reports none. Washington~D.C. has the highest advisory rate (85.5\%), indicating frequent detection of low-level risks without escalation.

\begin{table}[ht]
\centering
\caption{Per-city performance summary (Argoverse~2, n\,=\,1{,}000).}
\label{tab:cities}
\small
\setlength{\tabcolsep}{2pt}
\begin{minipage}{0.49\linewidth}
\centering
\begin{tabular}{%
>{\raggedright\arraybackslash}p{1.1in}
>{\centering\arraybackslash}p{0.35in}
>{\centering\arraybackslash}p{0.5in}
>{\centering\arraybackslash}p{0.7in}}
\hline
\textbf{City} & \textbf{n} & \textbf{Mean} & \textbf{Emerg.\,\%} \\
\hline
Miami      & 260 & 66.4 & 7.3\% \\
Austin     & 229 & 70.4 & 3.5\% \\
Pittsburgh & 205 & 67.3 & 5.4\% \\
\hline
\end{tabular}
\end{minipage}
\hfill
\begin{minipage}{0.49\linewidth}
\centering
\begin{tabular}{%
>{\raggedright\arraybackslash}p{1.1in}
>{\centering\arraybackslash}p{0.235in}
>{\centering\arraybackslash}p{0.5in}
>{\centering\arraybackslash}p{0.7in}}
\hline
\textbf{City} & \textbf{n} & \textbf{Mean} & \textbf{Emerg.\,\%} \\
\hline
Washington D.C. & 124 & 68.3 & 5.6\% \\
Dearborn        & 122 & 71.2 & 0.8\% \\
Palo Alto       &  60 & 71.7 & 0.0\% \\
\hline
\end{tabular}
\end{minipage}
\end{table}

\subsection{Intervention Tier Analysis}
\label{subsec:results_tiers}

\begin{figure}[ht]
\centering
\includegraphics[width=0.75\linewidth]{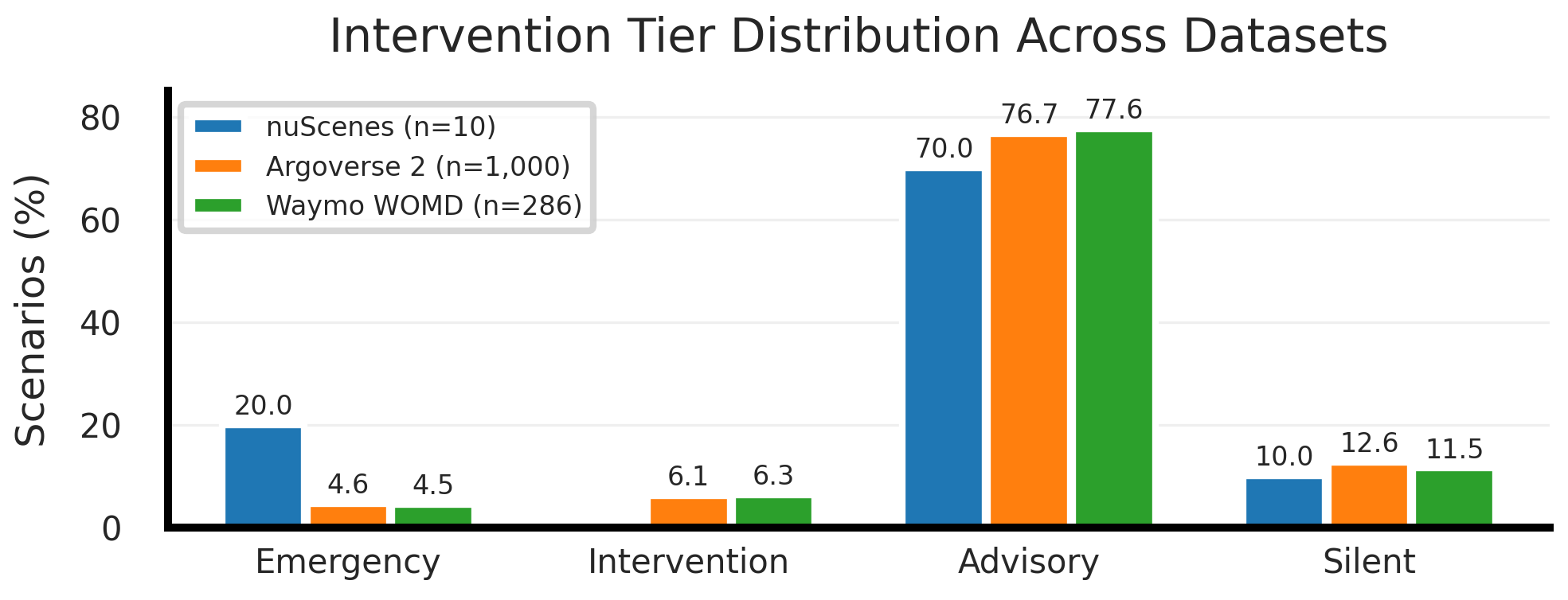}
\caption{Intervention tier distribution across all three evaluation datasets. The advisory tier consistently dominates across datasets (70--78\%), while the elevated nuScenes emergency rate (20\%) reflects adverse-condition sampling rather than model miscalibration.}
\label{fig:tiers}
\end{figure}

Figure~\ref{fig:tiers} presents the distribution of intervention tiers across all three datasets. The advisory tier is consistently dominant: 70\% on nuScenes, 76.7\% on Argoverse~2, and 77.6\% on Waymo. This consistent pattern across independently collected datasets, without retraining, confirms that PRISM's graduated response calibrates appropriately to the moderate-risk profile of structured urban driving.

The emergency tier represents 20\% of nuScenes scenes, compared to 4.6\% in Argoverse~2 and 4.5\% in Waymo. The higher emergency rate in nuScenes results from its adverse-condition sampling bias, not model miscalibration. There is no evidence of silent-tier over-suppression. Silent classifications require a score above 70, which PRISM assigns only when trajectory risk, VRU proximity, and environmental conditions are all simultaneously low.

The intervention tier (6.1\% AV2, 6.3\% Waymo) and silent tier (12.6\% AV2, 11.5\% Waymo) are equally well-matched across the two large datasets, reinforcing calibration consistency across all four tiers.


Waymo's traffic composition differs substantially from Argoverse~2, averaging 52.7 vehicles and 5.9 pedestrians per scene. This results in the highest mean trajectory risk among the three datasets (0.87), primarily due to dense multi-vehicle interactions. Despite this elevated kinematic signal, the VRU-dominant fusion weights ($w_v = 1.0$, $w_t = 0.5$ in Equation~(3)) reduce this impact, producing a mean safety score of 68.0 that closely matches Argoverse~2 (68.6) and a comparable risk profile (Table~\ref{tab:waymo}). These results indicate that the fusion design does not overstate risk in vehicle-dense, pedestrian-sparse scenarios.

\begin{table}[ht]
\centering
\caption{Waymo WOMD risk profile ($n$\,=\,286).}
\label{tab:waymo}
\small
\begin{tabular}{lrlr}
\hline
Mean safety score       & 68.0 & Avg.\ vehicles/scene    & 52.7 \\
Mean trajectory risk    & 0.87 & Avg.\ pedestrians/scene &  5.9 \\
Mean VRU risk           & 0.06 & Near-miss rate          & 3.8\% \\
\hline
\end{tabular}
\end{table}

\subsection{VRU Proximity and Near-Miss Detection}
\label{subsec:results_vru}

\begin{figure}[ht]
\centering
\includegraphics[width=0.95\linewidth]{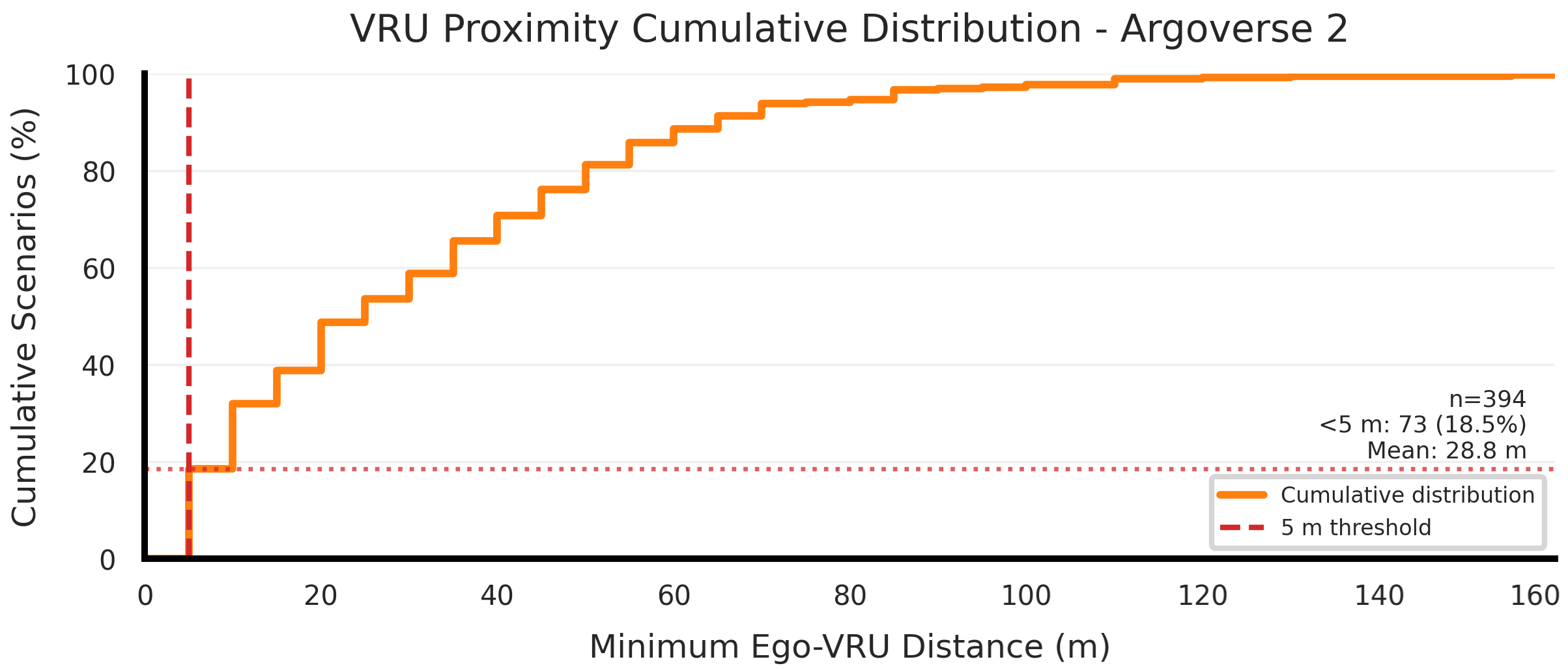}
\caption{Cumulative distribution of minimum ego-VRU distance across 394 Argoverse~2 scenarios with VRU present. The dashed red line marks the 5~m near-miss threshold; 18.5\% of VRU-present scenarios (73 of 394) fall within this range, all classified as intervention or emergency tier.}
\label{fig:vru}
\end{figure}

PRISM detects VRU near-miss events by applying the Social Force Model with a conservative collision-course threshold. Across nuScenes, only scene-0061 triggered a confirmed near-miss, involving 60 pedestrians at a minimum ego-VRU distance of 2.4 m, resulting in a VRU risk of 0.72 and an emergency-tier classification. The average minimum VRU distance in nuScenes was 16.2 m, with three other scenes below 5.5~m (vru\_risk $>$ 0.09). The night rain scene (scene-1094, 55 pedestrians, 3.1 m) had the second-highest VRU risk (0.43), confirming that the environmental multiplier amplifies proximity risk under adverse conditions.

On Waymo, 11 of 286 scenarios (3.8\%) triggered near-miss detections, with a mean VRU risk of 0.063, reflecting the lower pedestrian density (average 5.9 per scene compared to 28.6 in nuScenes). Figure~\ref{fig:vru} shows the cumulative ego-VRU proximity distribution for Argoverse~2, 18.5\% of VRU-present scenarios (73 of 394) had a minimum distance below 5~m, all of which received intervention or emergency tier classifications.

\subsection{Adverse Condition Sensitivity}
\label{subsec:results_adverse}

nuScenes is the only dataset with labeled adverse conditions, comprising three night scenes and one simultaneous night-and-rain scene. Table~\ref{tab:adverse} summarizes mean safety scores and environmental multipliers by condition group.

\begin{table}[ht]
\centering
\caption{PRISM output by environmental condition (nuScenes, n=10).}
\label{tab:adverse}
\small
\begin{tabular}{lrrrr}
\hline
\textbf{Condition} & \textbf{n} & \textbf{Mean Score} & \textbf{Env. Multiplier} & \textbf{Dominant Tier} \\
\hline
Clear daytime   & 7 & 66.2 & 0.97 & Advisory (6/7) \\
Night only      & 2 & 55.7 & 1.33 & Advisory (2/2) \\
Night + Rain    & 1 & 23.3 & 1.35 & Emergency      \\
\hline
\end{tabular}
\end{table}

Night conditions increase the environmental multiplier from 0.97 to 1.33, lowering mean safety scores by 10.5 points (from 66.2 to 55.7) compared to clear daytime. In the night-and-rain scene (scene-1094), the score drops to 23.3 due to a multiplier of 1.35 combined with 55 pedestrians at 3.1~m. The combination of adverse lighting, precipitation, and high VRU density places this scene firmly in the emergency tier, demonstrating that PRISM's multiplicative environmental layer effectively amplifies dynamic risk signals under compound adverse conditions.

The clear daytime emergency scene (scene-0061, score 35.8) shows that emergency-tier classification can result solely from high VRU density (60 pedestrians at 2.4~m, near-miss confirmed), regardless of environmental factors. The only silent-tier scene (scene-0757, score 85.6) had low VRU density and clear conditions, confirming that PRISM does not over-escalate risk in genuinely safe situations.

\subsection{SHAP Feature Attribution}
\label{subsec:results_shap}

\begin{figure}[ht]
\centering
\includegraphics[width=0.9\linewidth]{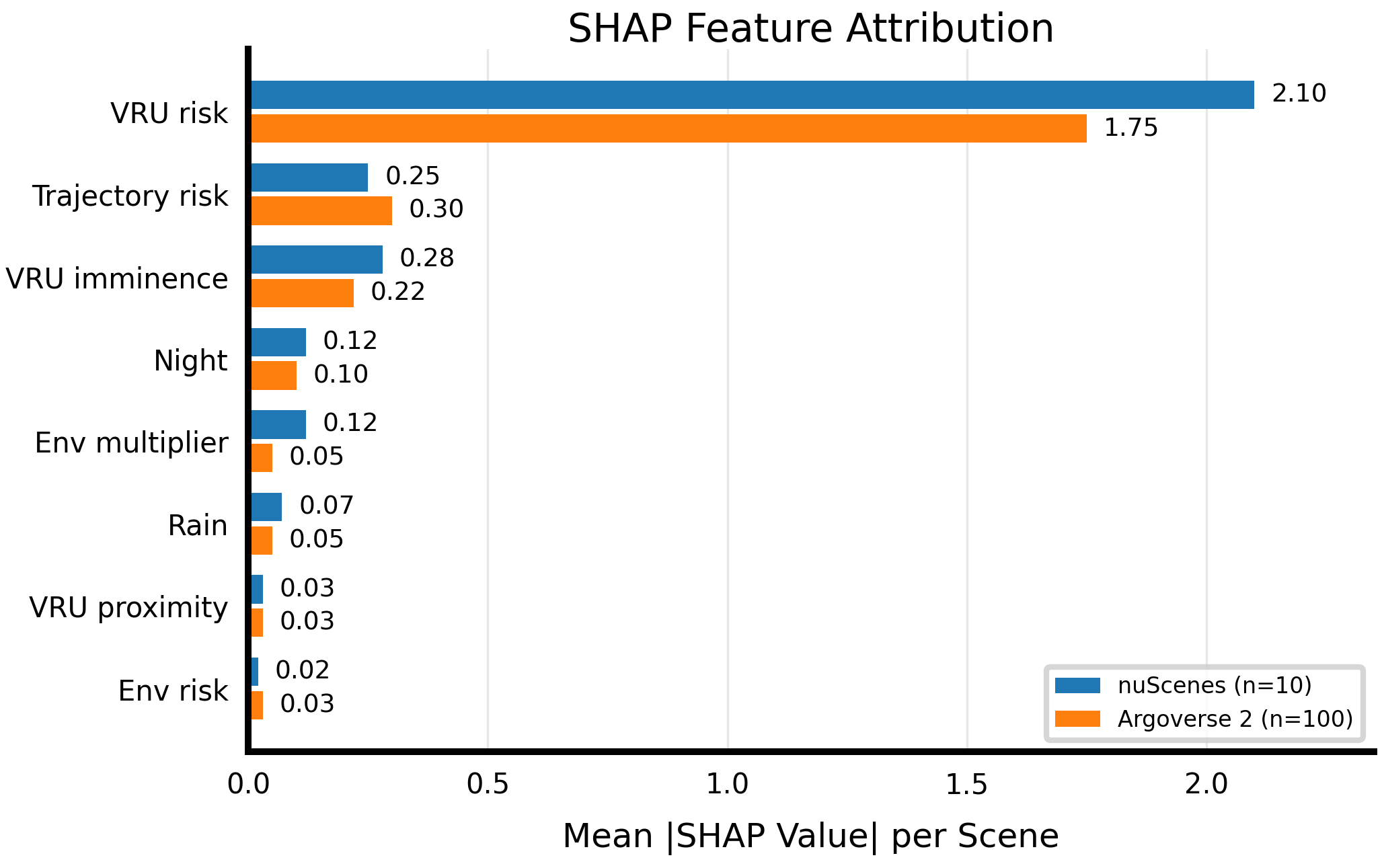}
\caption{Mean $|\text{SHAP}|$ per scene for the PRISM RL agent on nuScenes (n=10) and Argoverse~2 (n=100). VRU risk is the dominant driver in both datasets; environmental features contribute consistently but at a substantially lower magnitude.}
\label{fig:shap}
\end{figure}

Figure~\ref{fig:shap} presents the top SHAP feature attributions for the PRISM RL agent on nuScenes and a 100-scenario Argoverse~2 sample.

\begin{itemize}
\item \textbf{nuScenes:} VRU risk is the primary driver (mean $|\text{SHAP}| = 2.10$ per scene), followed by VRU imminence (0.28) and trajectory risk (0.25). Environmental features (night, env multiplier, rain) contribute but at an order of magnitude less.
\item \textbf{Argoverse~2:} VRU risk remains dominant (1.75 per scene), with trajectory risk (0.30) ranking second and VRU imminence third (0.22). Environmental features follow the same secondary pattern.
\end{itemize}

The feature ranking is consistent across both independently collected datasets without retraining, confirming that the learned decision policy generalizes across driving environments.

\subsection{Ablation Study: Feature Group Contribution}
\label{subsec:results_ablation}

\begin{figure}[ht]
\centering
\begin{minipage}[t]{0.42\linewidth}
    \vspace{0pt}
    \centering
\footnotesize
\phantomsection\label{tab:ablation}
\begin{tabular}{p{1.8cm}p{0.6cm}p{1.2cm}p{1cm}}
\hline
\textbf{Config} & \textbf{Mean Tier} & \textbf{Escalated} & \textbf{Emerg.} \\
\hline
ENV only         & 0.0 & 0/10 & 0 \\
ENV + Trajectory & 0.8 & 8/10 & 0 \\
ENV + VRU        & 0.4 & 2/10 & 0 \\
Full PRISM       & 1.3 & 9/10 & 2 \\
\hline
\end{tabular}
\end{minipage}
\hfill
\begin{minipage}[t]{0.56\linewidth}
    \vspace{0pt}
    \centering
    \includegraphics[width=\linewidth]{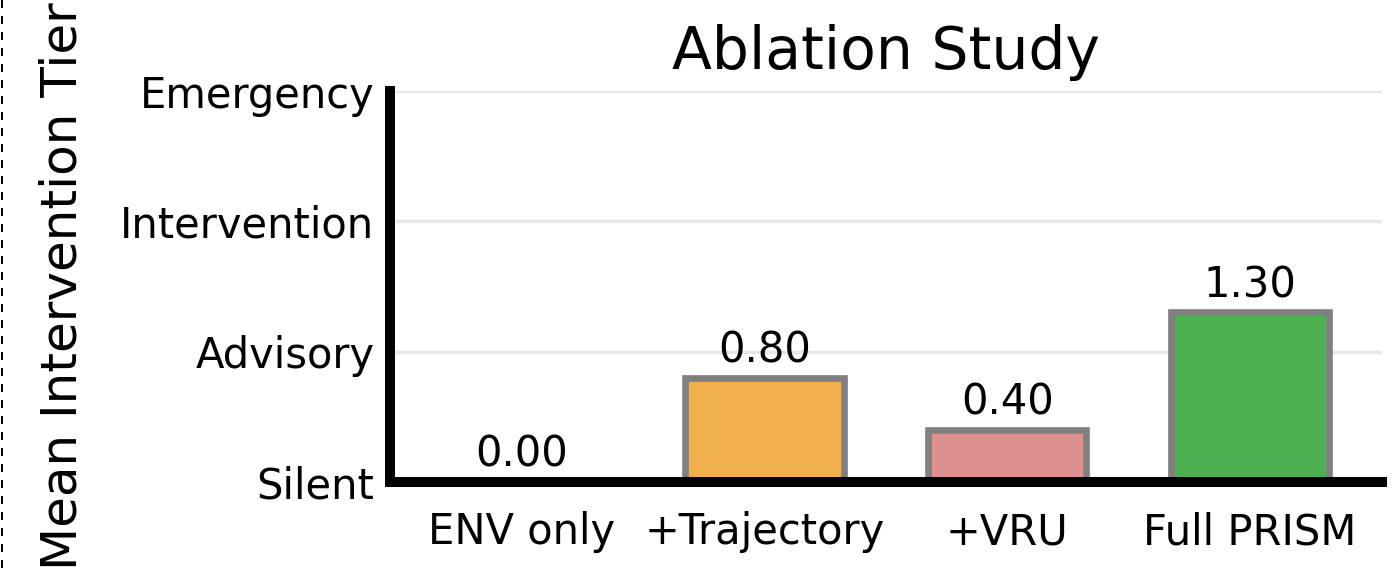}
\end{minipage}
\caption{Ablation study (nuScenes, n=10): Mean intervention tier by configuration, as a summary table (left) and bar chart (right). Tier encoding: 0 = silent, 1 = advisory, 2 = intervention, 3 = emergency. Only full PRISM with the RL agent reaches the emergency tier.}
\label{fig:ablation}
\end{figure}

Figure~\ref{fig:ablation} presents the marginal contribution of each model component, evaluated on nuScenes (n=10). With only the Phase~1 environmental model, all ten scenes are classified as silent. The RF bridge accurately captures environmental context but does not respond to dynamic trajectory or VRU signals, reflecting the known limitation of the Phase~1 system \citep{safedriveriq}.

Including the trajectory model escalates 8 of 10 scenes to advisory, capturing kinematic risk from aggressive or high-speed maneuvers. Adding the VRU model independently escalates the two highest-proximity scenes (scene-0061 and scene-1094) directly to intervention, confirming that the Social Force Model identifies high-risk ego–VRU encounters even when trajectory risk is moderate. Neither component alone reaches the emergency tier. The full PRISM system, with the RL agent synthesizing all three signals, correctly classifies both scenes as emergencies and maintains advisory for the remaining 7 kinematically active scenes.

\subsection{Tier Boundary Sensitivity}
\label{subsec:results_sensitivity}

To assess robustness of tier boundaries, each composite-score boundary has been adjusted by $\pm5$ points and recalculated tier distributions for all 1{,}000 Argoverse~2 scenarios (Table~\ref{tab:sensitivity}).

\begin{table}[ht]
\centering
\caption{Tier distribution under $\pm5$-point boundary perturbation, Argoverse~2.}
\label{tab:sensitivity}
\small
\begin{tabular}{lrrrr}
\hline
\textbf{Boundaries} & \textbf{Emer.\,\%} & \textbf{Interv.\,\%} & \textbf{Advisory\,\%} & \textbf{Silent\,\%} \\
\hline
Baseline         & 4.5 &  6.1 & 76.7 & 12.7 \\
Shift $+5$ pts   & 6.7 &  5.9 & 85.3 &  2.1 \\
Shift $-5$ pts   & 3.0 &  5.8 & 36.5 & 54.7 \\
\hline
\end{tabular}
\end{table}

The combined emergency and intervention escalation rate varies from 8.8\% (Shift~$-5$) to 12.6\% (Shift~$+5$), a range of less than 4 percentage points, confirming that safety-critical tier assignments are robust to minor calibration errors. The advisory and silent tiers are more sensitive near the 76-point threshold, but since both indicate non-escalated safe conditions, this sensitivity does not impact safety-critical decisions.

\subsection{Computational Latency}
\label{subsec:results_latency}

\begin{figure}[ht]
\centering
\includegraphics[width=0.75\linewidth]{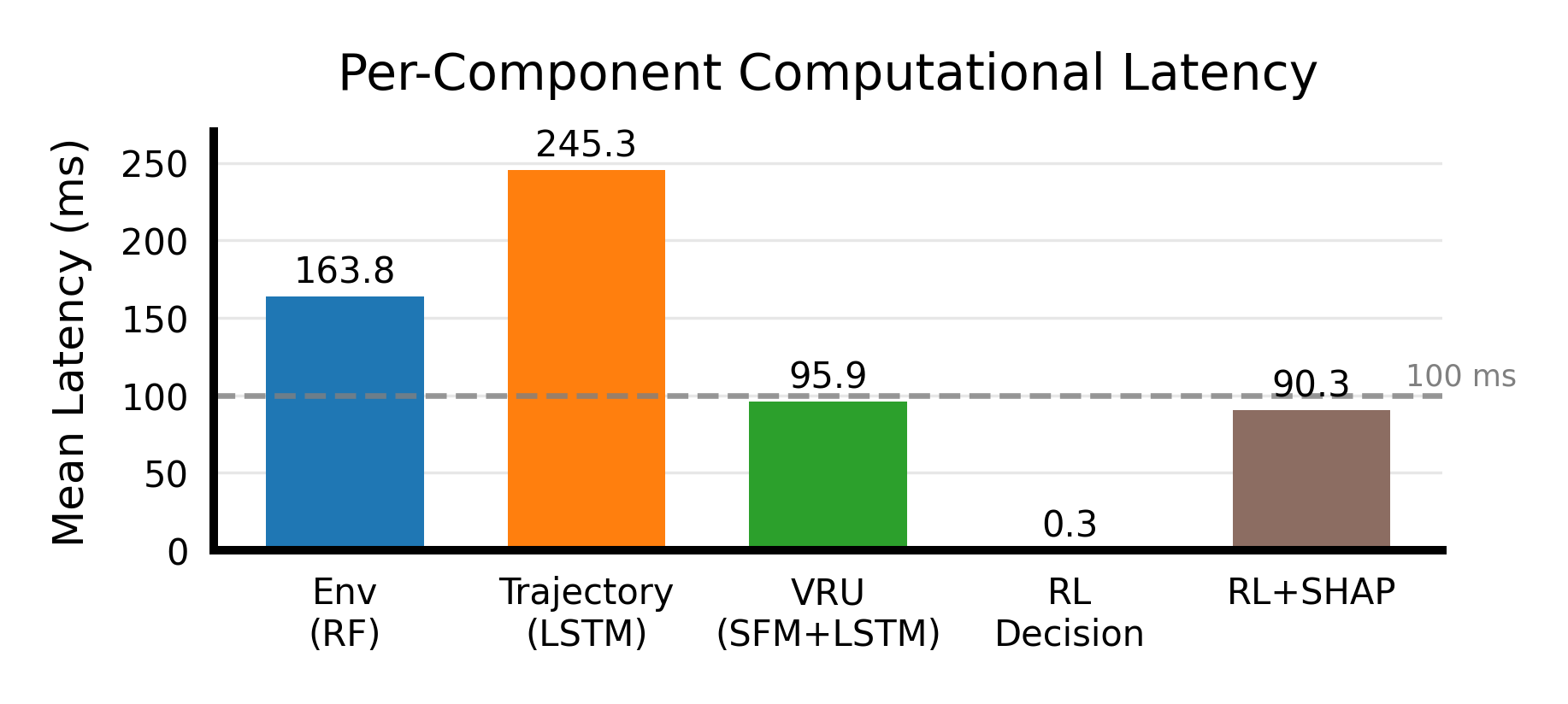}
\caption{Per-component mean computational latency on CPU (n=20 samples). The trajectory LSTM and environmental RF bridge account for 68\% of total processing time. The RL decision step contributes less than 1~ms.}
\label{fig:latency}
\end{figure}

Figure~\ref{fig:latency} profiles the per-component mean latency of PRISM on CPU (n=20 samples). The trajectory LSTM is the dominant cost at 245.3~ms, followed by the environmental RF bridge at 163.8~ms. The VRU module (SFM + LSTM) contributes 95.9~ms, and SHAP attribution adds 90.3~ms when explanations are requested. The RL decision step itself is negligible at 0.3~ms. Total end-to-end latency is approximately 596~ms on CPU, which is consistent with an offline post-processing role rather than hard real-time control. GPU deployment or ONNX-optimized inference is expected to reduce this by an order of magnitude, as discussed in Section~\ref{sec:future}.

\subsection{Risk Factor Co-occurrence}
\label{subsec:results_cooccurrence}

Both emergency-tier scenes in nuScenes involve two or more risk dimensions occurring simultaneously: scene-0061 (confirmed VRU near-miss and high trajectory risk, $r_{\text{traj}} = 0.55$) and scene-1094 (VRU near-miss, high trajectory risk $r_{\text{traj}} = 0.83$, and adverse conditions from night and rain). Scenes with only a single elevated risk dimension remained at the advisory tier, consistent with Phase~1 findings that compound risk factors lead to the most severe outcomes \citep{safedriveriq}. The RL agent's ability to synthesize signals across three independent model streams is precisely what enables detection of these compound events, which no single model alone would escalate to an emergency.
\section{Limitations}
\label{sec:limitations}

\textbf{Data and Evaluation.} PRISM's evaluation has several limitations. The nuScenes subset comprises only ten scenes, which limits the statistical power of conclusions about adverse conditions. For example, the observed increase in emergency rates under night-and-rain conditions is directionally consistent but not statistically generalizable. Additionally, none of the datasets used include annotated near-miss ground truth. As a result, VRU detection performance is reported as detection rate against a threshold-based proxy, rather than as precision and recall against labeled events.

\textbf{Model and Deployment.} The intervention tier boundaries are derived from domain knowledge rather than learned from labeled intervention data, and their optimality has not been independently validated. The RL agent was trained exclusively on nuScenes. While cross-dataset tier consistency is encouraging, it does not guarantee formal generalization. Finally, the end-to-end latency of approximately 596ms on the CPU prevents hard real-time deployment. GPU or ONNX-optimized inference is required for on-vehicle integration.
\section{Future Directions}
\label{sec:future}

\textbf{Latency Optimization.} GPU deployment and ONNX-optimized inference are immediate priorities for on-vehicle integration. The trajectory LSTM and environmental RF bridge together account for 68\% of total latency and are the main targets for model compression and quantization.

\textbf{Learned Tier Calibration.} The current intervention tier boundaries are domain-knowledge-derived. A data-driven calibration using labeled intervention logs from naturalistic driving studies would replace fixed thresholds and improve sensitivity at tier boundaries, particularly between the advisory and intervention bands.

\textbf{RL Generalization.} Extending RL agent training to Argoverse 2 and Waymo scenarios would provide a formal generalization guarantee beyond the cross-dataset consistency observed in this study. Training on multiple datasets may also improve robustness to distribution shifts across sensor modalities and geographies.

\textbf{Ground Truth Evaluation.} Collaborating with fleet operators or simulation environments to obtain annotated near-miss labels would enable precision-recall evaluation of the VRU detector, replacing the current threshold-based proxy metric.

\textbf{V2X and Infrastructure Integration.} Incorporating vehicle-to-infrastructure (V2X) signals and roadside sensor feeds would extend PRISM beyond ego-vehicle perception. This integration would enable proactive warnings for occluded VRUs and intersection conflicts not visible to onboard sensors.

\textbf{V2X and Federated Learning}. Beyond infrastructure integration, a federated learning extension would enable PRISM to learn continuously from distributed fleet deployments while keeping sensitive trip data decentralized. This approach supports privacy and enables large-scale, real-world validation across various geographies and vehicle types.
\section{Conclusion}
\label{sec:conclusion}

This paper introduced PRISM, an agentic multi-model architecture for proactive safety assessment in autonomous transportation systems. By fusing three parallel risk models (environmental, trajectory kinematic, and VRU interaction) through a DQN reinforcement learning agent with SHAP explainability, this approach advances the Phase~1 SafeDriver-IQ foundation \citep{safedriveriq} from static crash-probability scoring to dynamic, scene-aware intervention decisions.

Evaluated across 1,296 scenarios from three publicly available autonomous driving datasets without dataset-specific retraining, PRISM achieved a mean safety score of 68/100, with 77.6\% of scenarios correctly classified as advisory under normal urban driving conditions. The emergency tier was triggered in 4.5–4.6\% of structured-environment scenarios and increased to 20\% under adverse conditions, reflecting the system's multiplicative environmental risk layer. SHAP attribution confirmed VRU risk as the primary decision driver. All emergency-tier classifications involved two or more active risk dimensions, consistent with Phase~1 findings on compound risk. The end-to-end latency of 596~ms on CPU highlights a clear optimization target for real-time deployment.

PRISM demonstrates that a modular, interpretable, and dataset-agnostic agentic architecture can generalize proactive safety assessment across diverse autonomous driving environments. It provides a deployable foundation for graduated intervention in next-generation transportation infrastructure.

\bibliography{References}
\end{document}